\pdfoutput=1

\documentclass[journal, 10pt, draftclsnofoot, onecolumn]{IEEEtran}
\usepackage{verbatim}
\ifCLASSINFOpdf
  \usepackage[pdftex]{graphicx}
\else
  \usepackage[dvips]{graphicx}
\fi
\usepackage[cmex10]{amsmath}
\ifCLASSOPTIONcompsoc
 \usepackage[caption=false,font=normalsize,labelfont=sf,textfont=sf]{subfig}
\else
 \usepackage[caption=false,font=footnotesize]{subfig}
\fi
\begin{document}
%
\title{General Geometric Fluctuation Modeling for Device Variability Analysis}
%
%
%

\author{
 Bo~Fu, Seonghoon~Jin, Woosung~Choi, Keun-Ho~Lee, Young-Kwan~Park  
\thanks{B. Fu, S. Jin and W. Choi are with the Device Lab, Samsung Semiconductor, Inc., 
USA, e-mail: bo.f@ssi.samsung.com.} 
\thanks{K.H.~Lee, Y.K.~Park are with Semiconductor R\&D Center, Samsung Electronics Co., Ltd., Korea.}
}

\maketitle

\begin{abstract}
The authors propose a new modeling approach based on the impedance field method (IFM) to analyze the general geometric variations in device simulations. Compared with the direct modeling of multiple variational devices, the proposed geometric variation (GV) model shows a better efficiency thanks to its IFM based nature. Compared with the existing random geometric fluctuation (RGF) model where the noise sources are limited to the interfaces, the present GV model provides better accuracy and wider application areas as it transforms the geometric variation into global mesh deformation and computes the noise sources induced by the geometric variation in the whole simulation domain. GV model also provides great insights into the device by providing the effective noise sources, equation-wise contributions, and sensitivity maps that are useful for device characterization and optimization.
\end{abstract}


%
\IEEEpeerreviewmaketitle

\section{Introduction}
%
%
%
%

\IEEEPARstart {D}{evice} variability presents a growing challenge as we push forward Moore's Law. Various physical variability sources such as random dopant fluctuations (RDF), random workfunction fluctuations (RWF), and random geometric fluctuations (RGF) are becoming important as devices get smaller~\cite{4339737, 4418975, 5361404, 6131494, 6579670}. Among these variability sources, geometric variation becomes more significant as the uncertainty in the fabrication process such as lithography, etching, and deposition cannot be scaled with the advancing technology nodes. For example, line edge roughness (LER) caused by chemical inhomogeneity in the lithography and etching process is in the order of several nanometers and is not expected to decrease as the device size shrinks~\cite{ban2010electrical}. Geometric variation induced device fluctuations in threshold voltage, drive current, subthreshold swing, and drain-induced barrier lowering (DIBL) exhibit significant challenges for future VLSI circuit technology development.

In terms of the state of the art geometric variation analysis, people either apply direct modeling with sufficient device samples~\cite{6131494, 1210771} or RGF model~\cite{guide201409synopsys}. The former is reliable, however it greatly suffers in terms of efficiency. Thanks to the impedance field method (IFM)~\cite{658840}, the latter can provide efficient solutions if variations are small such that they are in the linear response range. However, since the RGF model constrains the noise source to be locally at the interface, it fails to consider nonlocal effects in geometric fluctuations. It is worth noting that the recent improvement of RGF model was made to capture the semiconductor body thickness variations. However, the accuracy is not guaranteed, and it relies on the calibration of parameters~\cite{guide201409synopsys}. The purpose of this study is to propose a rigorous IFM based model named GV (geometric variations), calculating general geometry variations in an efficient and accurate manner.

The paper is organized as follows: RGF model's limitations will be explored first regarding two geometric variation types: interface displacement and body deformation. Then we explain in detail the formulation of our GV method. We will then validate our model's effectiveness by comparing with numerical experiments on different geometric variation cases in the two dimensional device simulations. For a three dimensional application, we demonstrate the GV model's usage on a FinFET structure in which we have considered channel length, body thickness, and fin height variations. During the process of the FinFET GV simulations, new physical insights have been obtained such as the effective noise sources, equation-wise contributions, and sensitivity maps. Finally, we make a summary and conclusion.

\section{RGF Limitations}
We have tested Sentaurus\textsuperscript{TM} Device RGF model~\cite{guide201409synopsys} for interface displacement variability calculations. However, inconsistent results are observed in comparison with the designed numerical experiments. By checking the calculated effective Green's functions for RGF and examining the noise source described in~\cite{guide201409synopsys}, we conclude that the RGF model is incapable of modeling general geometric variations.

\subsection{Interface Displacement}
RGF model accounts for the interface displacement and its low frequency noise source is given as~\cite{guide201409synopsys}:
\begin{align}
\label{eq:RGF_model}
K(\vec{r_1},\vec{r_2}){} ={} &\hat{\vec{n}}(\vec{r_1}) \cdot \hat{\vec{n}} (\vec{r_2})\exp(-\frac{ ||\vec{r_1}-\vec{r_2}||^2}{\lambda^2}) \nonumber \\
&[a_{\text{iso}} (\vec{r_1})+|\vec{a} (\vec{r_1}) \cdot \hat{\vec{n}} (\vec{r_1})|] \nonumber \\
&[a_{\text{iso}} (\vec{r_2})+|\vec{a} (\vec{r_2}) \cdot \hat{\vec{n}} (\vec{r_2})|],
\end{align}
where $\vec{r_1}$ and $\vec{r_2}$ are positions on the interface, $\hat{\vec{n}}(\vec{r})$ is the local interface normal in position $\vec{r}$, $a_{\text{iso}}$ the isotropic correlation amplitude, $\vec{a} (\vec{r})$ the vectorial correlation amplitude, and $\lambda$ is the correlation length.

\begin{figure}[!t]
  \centering
  \includegraphics[width=3.0in]{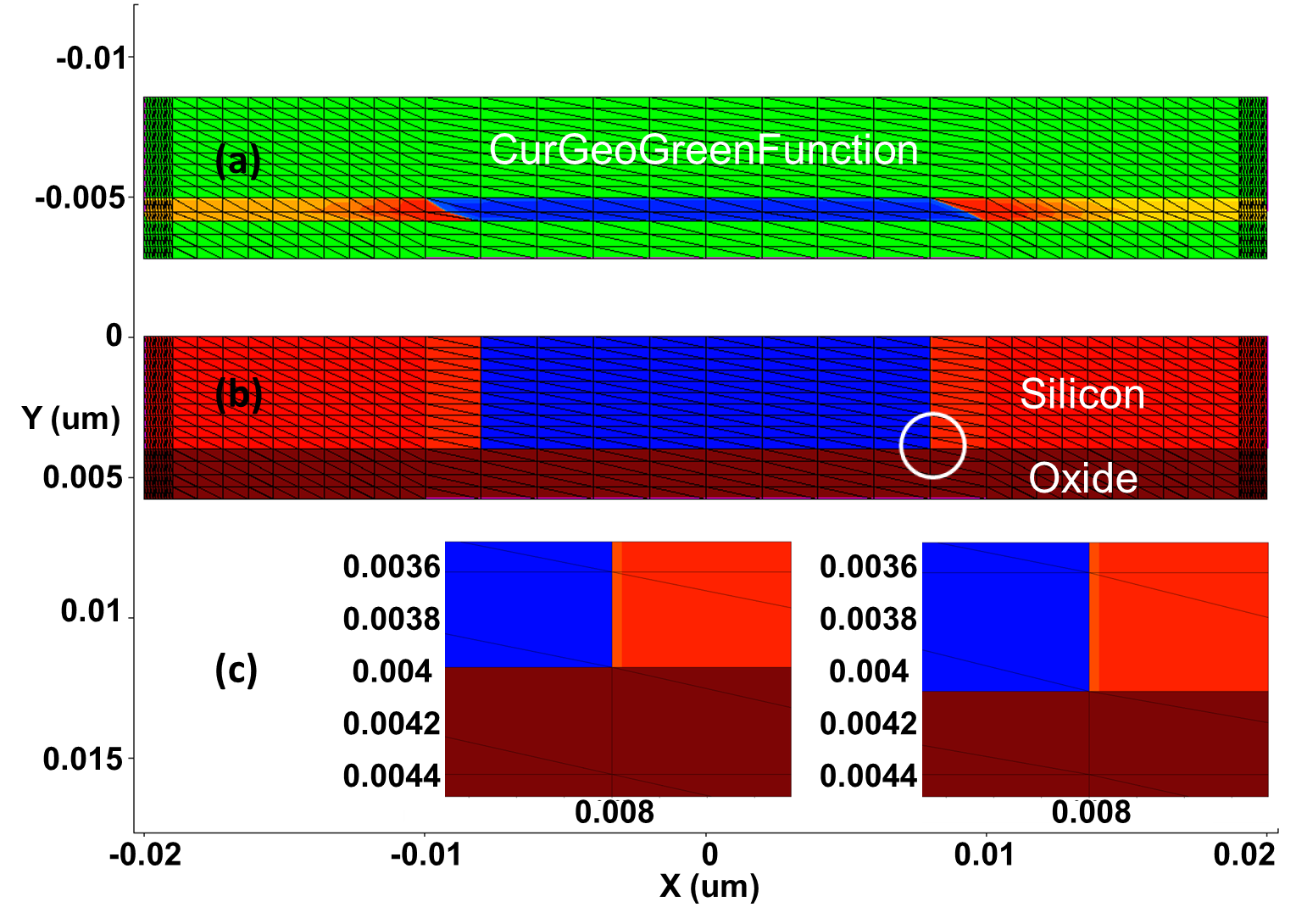}
  \caption{(a) Current Green's function for geometric variation, which only has non-zero values at the interface. (b) The nominal device, whose sizes are as shown in the X-Y coordinates. (c) An illustration of interface displacement. Whole silicon/oxide interface is shifted from $y=4$ nm to $y=4.1$ nm. Only the white circle area in (b) is zoomed for the nominal device (left) and variational device (right).}
  \label{fig:RGF_illus}
\end{figure}

To test the effectiveness of the model, a simple MOSFET structure as shown in Fig.~\ref{fig:RGF_illus}(b) is designed. The device is composed of only silicon and oxide, with source/drain donor concentration of $3.0\times10^{20}$ cm$^{-3}$ and acceptor concentration of $3.0\times10^{16}$ cm$^{-3}$ in all silicon region. We applied the RGF model on the ``silicon/oxide'' interface that is normal to the Y axis. By choosing a sufficiently large correlation length $\lambda$ = 1 m, the exponential term in (\ref{eq:RGF_model}) vanishes. Physically it is equivalent to shifting the whole interface simultaneously. Further by setting $\vec{a} (\vec{r})$ = (0, 0.1 nm, 0) and $a_{\text{iso}}$ = 0, the displacement is fixed to be 0.1 nm along the Y-axis direction. Accordingly, we can create a variational device as shown in the right of Fig.~\ref{fig:RGF_illus}(c) for the numerical experiment, in which the interface is moved from silicon to oxide by 0.1 nm. Numerical experimental results are obtained by calculating the difference of drain currents between the interface-shifted device and the nominal device.

\begin{figure}[!t]
  \centering
  \includegraphics[width=3.0in]{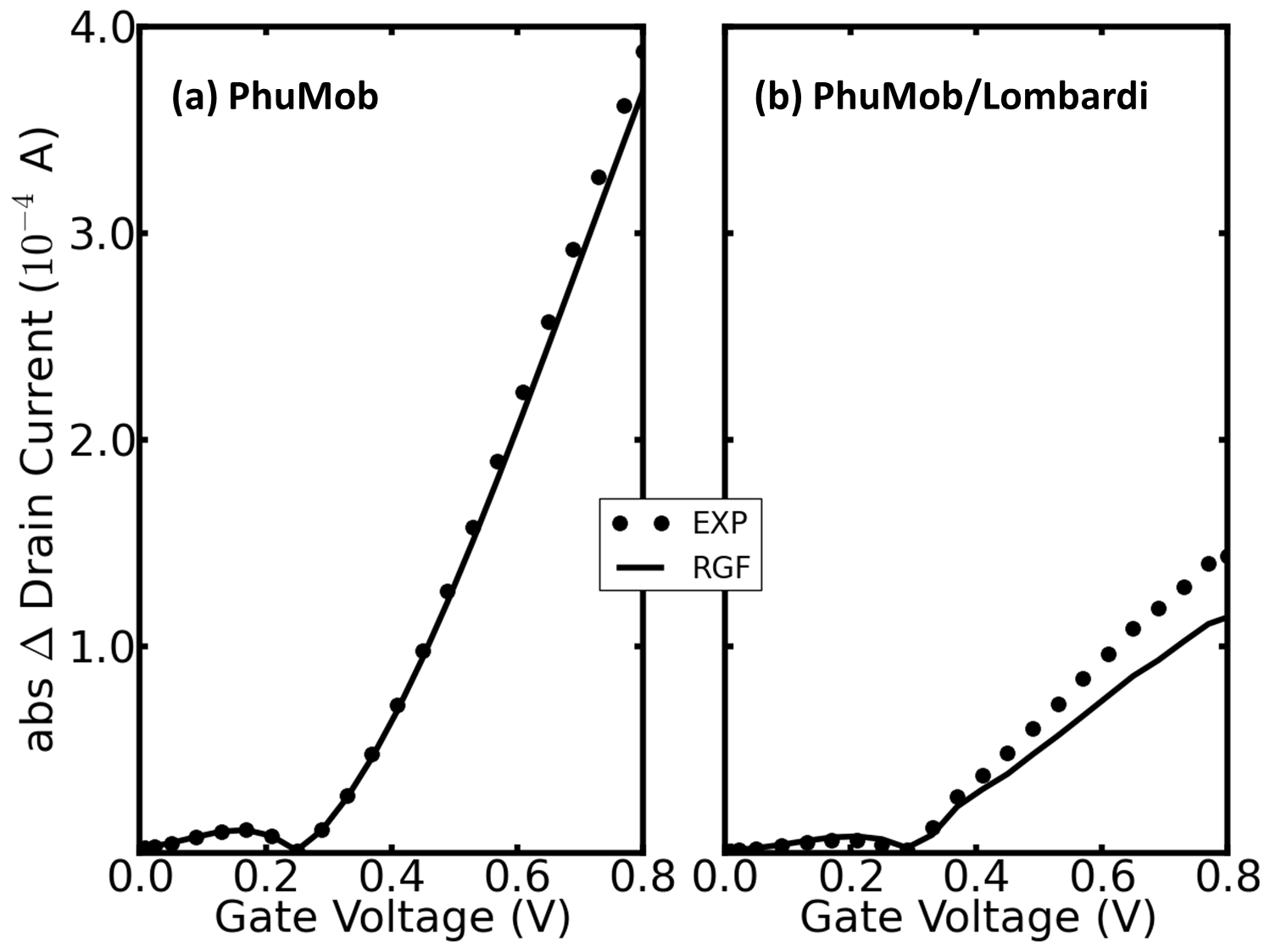}
  \caption{Results comparison of Sentaurus\textsuperscript{TM} Device RGF model with designed numerical experiments as described in Fig.~\ref{fig:RGF_illus}. Philips unified mobility model is applied in (a). In addition to the Philips unified mobility, enhanced Lombardi model is added in (b).}
  \label{fig:RGF_lomb}
\end{figure}

Noise-like IFM for RGF calculates the autocorrelation noise current spectral densities (SI). Giving our noise source $K(\vec{r_1},\vec{r_2})=(0.1~\text{nm})^2$ along the Y-axis, we can compare the square root of SI with the absolute values of numerical experimental results regarding drain current variations, as shown in Fig.~\ref{fig:RGF_lomb}. Philips unified mobility model~\cite{klaassen1992unified1} is used in Fig.~\ref{fig:RGF_lomb}(a) and enhanced Lombardi model~\cite{lombardi1988physically, darwish1997improved} is added in Fig.~\ref{fig:RGF_lomb}(b). Obvious discrepancy exists in Fig.~\ref{fig:RGF_lomb}(b) because the Lombardi model accounts for mobility degradation near the semiconductor/insulator interface, while RGF only captures the noise source locally at the interface as pointed out by (\ref{eq:RGF_model}) and demonstrated by the effective Green's function in Fig.~\ref{fig:RGF_illus}(a) which exhibits non-zero values only at the interface. Specifically, in the presence of Lombardi model, the low field mobility is given by~\cite{lombardi1988physically, darwish1997improved}:
\begin{equation}
\label{eq:mat_lomb}
\frac{1}{\mu} =\frac{1}{\mu_{\text{b}}} + D[\frac{1}{\mu_{\text{ac}}(F_{\perp})} + \frac{1}{\mu_{\text{sr}}(F_{\perp})}],
\end{equation}
where $\mu_{\text{b}}$ is the bulk mobility, $\mu_{\text{ac}}$ and $\mu_{\text{sr}}$ are the surface acoustic phonon mobility and surface roughness mobility both depending on the transverse electric field normal to the interface $F_{\perp}$, the damping factor is defined as~\cite{guide201409synopsys}:
\begin{equation}
\label{eq:mat_lomb_D}
D = \exp(-\frac{l}{l_{\text{crit}}}),
\end{equation}
where $l_{\text{crit}}$ is a fitting parameter with a default value of 10 nm, and $l$ is the distance from the interface. Interface displacement changes the $l$ and $F_{\perp}$ values of the nodes that are not at the interface, and hence their mobilities. This effect is not considered in the RGF model.

Similarly, RGF model fails when thin-layer mobility model~\cite{reggiani2007low} is applied. For high-field saturation model~\cite{shapiro1967carrier, canali1975electron}, the driving force of the node not at the interface can be also changed as it is proportional to the gradient of scalar variables. By considering just the noise source locally at the interface, RGF is also prone to give inaccurate results.

\subsection{Body Deformation}
Besides interface displacement, more general geometric variation is body deformation. For the MOSFET in Fig.~\ref{fig:RGF_illus}(b), body deformation may include but is not limited to silicon width, oxide thickness, and channel length variations. The recent improvement of RGF model added three options of calculating the thickness variation~\cite{guide201409synopsys}. Among the three options, one is used to control the calculation performed on one side or both sides of the interface, and the other two are fitting parameters. The basic assumption of the RGF model remains the same, which only relies on the local interface information as noise source to predict terminal variations. Thickness variation could be introduced by properly setting interface displacement in the latest RGF model. However, RGF model requires users to calibrate the two fitting parameters, which demonstrates that the physics is not fully captured. It becomes worse if we want to apply the RGF model for length variation because there is no clear and reasonable interface to pick as the origin of the noise source.

\section{GV Methodology}

\subsection{GV Formulation}
For continuum device simulations, we solve the following three coupled equations:
\begin{itemize}
  \item Poisson's equation
\begin{equation}
  \label{eq:poissonEqu}
 \mathcal{F}_{\Phi}:=-\nabla \epsilon  \cdot \nabla \Phi - \rho = 0,
\end{equation}
where $\epsilon$ is the permittivity, $\Phi$ is the electrostatic potential and $\rho$ is the net charge.

\item Electron and hole continuity equations
\begin{equation}
  \label{eq:contEqu}
 \mathcal{F}_{J} := \nabla \cdot \vec{F} + R = 0,
\end{equation}
where $\vec{F}$ is the carrier (electron or hole) flux, and $R$ is the net recombination rate.

\item Electron and hole density gradient quantum potential equations~\cite{ancona1989quantum}. For the electron case
\begin{equation}
  \label{eq:quanEqu_den}
\mathcal{F}_{\Lambda_n} := -\frac{\gamma \hbar^2}{6m} \frac{\nabla^2\sqrt{n}}{\sqrt{n}} - \Lambda_n = 0,
\end{equation}
where $\gamma$ is a fitting parameter, $m$ is the density-of-states effective mass, and $\Lambda$ is the quantum potential. (\ref{eq:quanEqu_den}) can be also solved in terms of the quantum potential formulation~\cite{902727}.


\end{itemize}

A general integral form of each equation $\alpha$ in ($\Phi$, $J$, $\Lambda$) is given by:
\begin{equation}
\label{eq:pde_int}
  \mathcal{N}_\alpha := \oint_{\Gamma} \vec{F} \cdot \hat{n} d \gamma - \int_{\Omega} S d \sigma = 0,
\end{equation}
where $\vec{F}$ and $S$ are the flux and source terms respectively, $\Gamma$ is the boundary of control volume $\Omega$ and $\hat{n}$ is the normal vector. 

After obtaining the steady-state solution for the nominal device with mesh $M$, we can calculate the current Green's Function $G^i_{\alpha}$ for terminal $i$ and equation $\alpha$ using the generalized adjoint method~\cite{1083675, bonani2001noise}. Since voltage and current Green's functions are mutually transformable using the impedance and admittance matrices, we only focus on the current calculation. The calculation of terminal voltage variation follows in a similar way.

Suppose the geometric variation introduces a new mesh $M'$, whose node coordinates are different from $M$ but topologically $M'$ is equivalent to $M$, i.e. the number of nodes and their connectivity remain unchanged. $M'$ can be used to reassemble (\ref{eq:poissonEqu})-(\ref{eq:quanEqu_den}) through (\ref{eq:pde_int}) with the already available steady-state scalar solutions ($\phi$, $n$, $p$, $\Lambda_n$, $\Lambda_p$, etc.) from the nominal device. For equation $\alpha$, the difference between the RHS on $M'$ and the RHS on $M$, $\delta \mathcal{N}_{\alpha} =  \mathcal{N}_{\alpha}(M') - \mathcal{N}_{\alpha}(M)$, is ultimately the effective noise source due to the geometric variation. The terminal current change then can be calculated as follows:
\begin{equation}
\label{eq:noise_cal}
\delta I^i = \sum_{\alpha \in (\Phi, J, \Lambda)}G^i_{\alpha} \delta \mathcal{N}_{\alpha},
\end{equation}
summing the contribution from each equation.

\subsection{Effective Noise Source Calculation}

\begin{figure}[!t]
  \centering
  \includegraphics[width=3.0in]{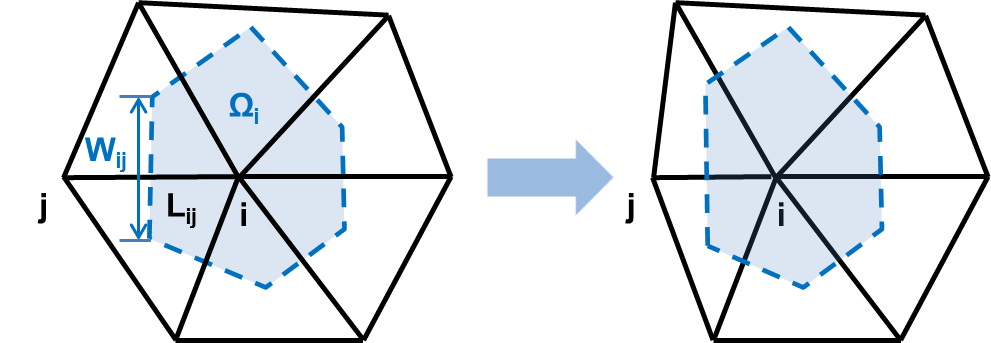}
  \caption{An illustration of finite volume method discretization in 2D. For nodes $i$ and $j$, $W_{ij}$ is the cross-section length and $L_{ij}$ is the side-length. $\Omega_{i}$ (dashed blue box) is the control volume of node $i$. Geometric variation moves node $j$ closer to node $i$.}
  \label{fig:fvm_illus}
\end{figure}
Finite volume method (FVM)~\cite{markowich1990semiconductor} can be used to solve (\ref{eq:pde_int}). A plot of FVM discretization is illustrated in Fig.~\ref{fig:fvm_illus}, in which box coefficient can be defined as $A_{ij}=W_{ij}/L_{ij}$, the cross-section to side-length ratio, and node volume $\Omega$ is the area of the dashed blue box. Intuitively, geometric variation changes only box coefficient $A$ and node volume $\Omega$. Under this assumption, the effective noise source from equation $\alpha$ can be obtained as:
\begin{equation}
\label{eq:noise_cal_linear}
\delta \mathcal{N}_{\alpha} = \frac{\partial \mathcal{N}_{\alpha}}{\partial A} \delta A + \frac{\partial \mathcal{N}_{\alpha}}{\partial \Omega} \delta \Omega 
\end{equation}
where $\delta A$ and $\delta \Omega$ are the box coefficient and volume changes, respectively. This is similar to the method by A. Gnudi, et al~\cite{gnudi1987sensitivity}. Instead of applying Green's function on each equation as in (\ref{eq:noise_cal}), their approach is to convert ($\delta \Omega$, $\delta A$) to ($\delta \phi$, $\delta n$, $\delta p$) to continue the variability calculations.

However, the assumption that only $\delta A$ and $\delta \Omega$ change is not always valid. Three additional consequences of geometric variations should be considered: 
\begin{itemize}
\item Shift of node j as shown in Fig.~\ref{fig:fvm_illus} changes all the distances between j and other nodes. If a semiconductor/insulator interface consists of node j, a different distance value affects the flux term at the other node. This component is critical to correctly simulate geometric dependence of, for example, the enhanced Lombardi model. 
\item Within the elements where node j belongs to, element gradient coefficients will be different, which alters element vectors such as electric field calculated from scalars. Geometric dependence of high field mobility usually requires this component. 
\item If node j happens to be on the internal boundary, a proper boundary condition should also be taken into account, which occurs for example when we want to simulate wave function penetration into insulator using (\ref{eq:quanEqu_den}) by applying an inhomogeneous Neumann boundary condition~\cite{902727}. 
\end{itemize}

In order to emphasize the importance of all components, we have identified and extracted them during reassembling the RHS of (\ref{eq:pde_int}) on the variational mesh $M'$. Detailed comparison of each component's contribution is shown in the application section.




\subsection{Geometric Variation Mesh Generation}
The present GV model requires that the variational mesh $M'$ and nominal device mesh $M$ share the same topology. For this purpose, we have created a strained mesh scheme to only change vertex coordinates. In three dimension, for vertex ($x_0$, $y_0$, $z_0$), the change of coordinate ($\delta x_0, \delta y_0, \delta z_0$) is:
\begin{equation}
\label{eq:strainMesh}
\begin{cases}
\delta x_0 = \Delta x \cdot f_{\text{XX}}(x_0) \cdot f_{\text{XY}}(y_0) \cdot f_{\text{XZ}}(z_0)  &\\
\delta y_0 = \Delta y \cdot f_{\text{YX}}(x_0) \cdot f_{\text{YY}}(y_0) \cdot f_{\text{YZ}}(z_0)  &\\
\delta z_0 = \Delta z \cdot f_{\text{ZX}}(x_0) \cdot f_{\text{ZY}}(y_0) \cdot f_{\text{ZZ}}(z_0)
\end{cases}
\end{equation}
where ($\Delta x$, $\Delta y$, $\Delta z$) are predefined amplitudes of geometric change along the (X, Y, Z) directions. A modulation function, for example $f_{\text{XY}}(y)$, is defined on the X-Y plane, describing the $y$'s influence on $x$. Consequently, the vertex ($x_0$, $y_0$, $z_0$) in nominal device mesh $M$ will be moved to ($x_0 + \delta x_0$, $y_0 + \delta y_0$, $z_0 + \delta z_0$) in the new mesh $M'$.

\begin{figure}[!t]
\centering
\subfloat[]{
  \includegraphics[width=1.0in]{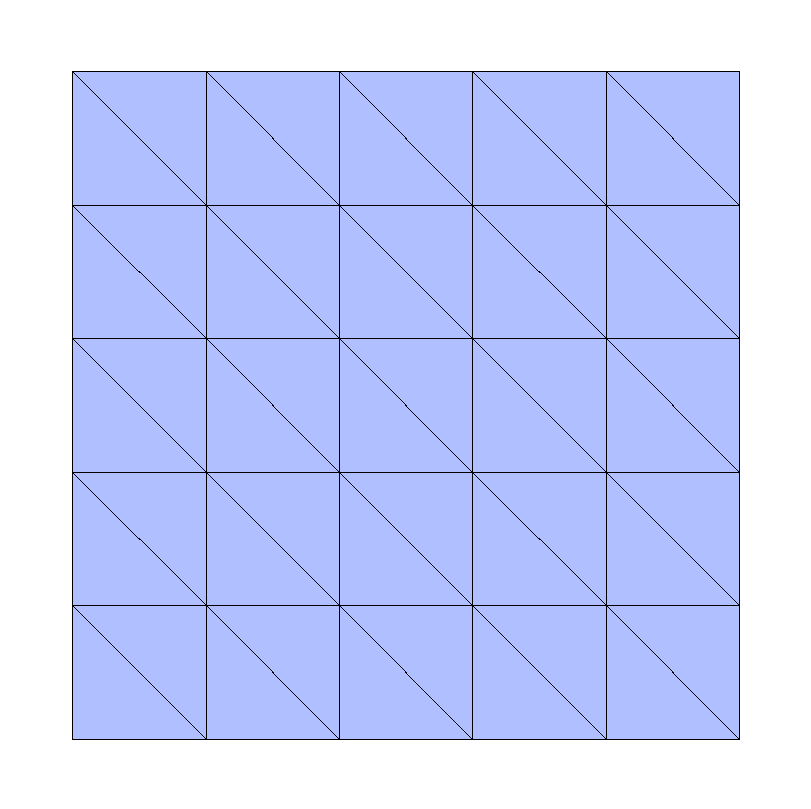}
  \label{fig:strainMesh_0}
}
\hfil
\subfloat[]{
  \includegraphics[width=1.0in]{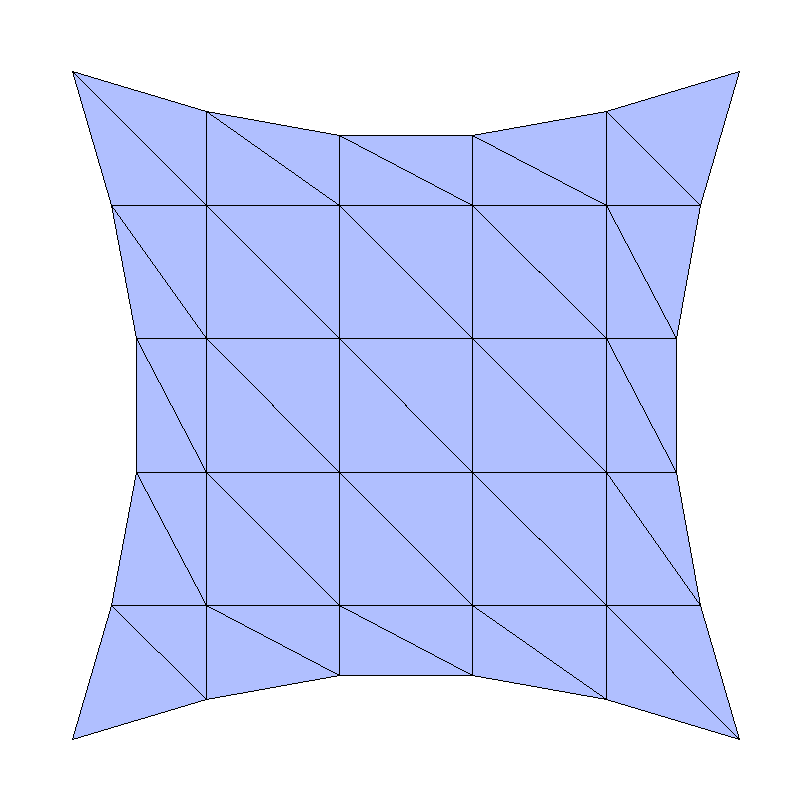}
  \label{fig:strainMesh_1}
}
\hfil
\subfloat[]{
  \includegraphics[width=1.0in]{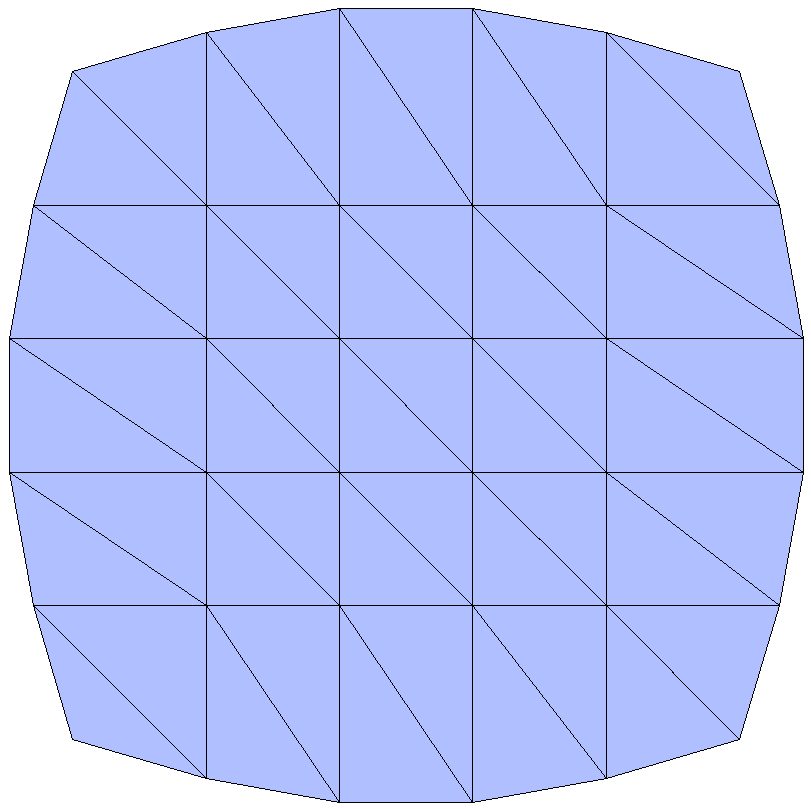}
  \label{fig:strainMesh_2}
}
\caption{A toy application of strained mesh scheme to mimic geometric variations: (a) a square is transformed to have (b) inward sine functions on four sides and (c) outward sine functions on four sides. All shapes have the same node connectivity but varied node coordinates.}
\label{fig:strainMesh}
\end{figure}

Fig.~\ref{fig:strainMesh} demonstrates a toy application of the strained mesh scheme: every side of a square (Fig.~\ref{fig:strainMesh}(a)) is modified by the inward sine function (Fig.~\ref{fig:strainMesh}(b)) or by the outward sine function (Fig.~\ref{fig:strainMesh}(c)). Although shapes change significantly from Fig.~\ref{fig:strainMesh}(a) to (b) and (c), their node connectivities remain the same. More practical applications of the strained mesh schema will be shown in the application section.

\section{Applications}
Using our in-house device simulation tool~\cite{7047005} which includes the GV model, we have performed the following device simulations. Regarding mobility models, we have employed Philips unified mobility~\cite{klaassen1992unified1}, enhanced Lombardi surface mobility~\cite{lombardi1988physically, darwish1997improved}, and high field saturation mobility~\cite{shapiro1967carrier, canali1975electron}. 

The effectiveness of our GV model can be validated by comparing with numerical experiments. Consider that the strained meshes created for GV analysis are variational devices that we can run direct simulations on. Drain current change $\Delta I_{\text{D}}$ induced by a geometric variation therefore can be obtained by comparing the $I_{\text{D}}\text{-}V_{\text{G}}$ characteristic of a variational device on  $M'$ and the nominal device on $M$, i.e. $\Delta I_{\text{D}} = I_{\text{D}}(M') -I_{\text{D}}(M)$, which should agree with the result calculated by the GV model. 

For the two dimensional device application, we mainly focus on the model validation by studying both interface displacement and body deformation. In the practical three dimensional FinFET example, based on comparison with numerical experiments, linear range of GV model will be examined. Contributions of effective noise source components besides box coefficients will be compared. We also look at the equation-wise contributions and the geometric sensitivity maps.


\subsection{Two Dimensional DG-FET}
\begin{figure}[!t]
  \centering
  \includegraphics[width=3.0in]{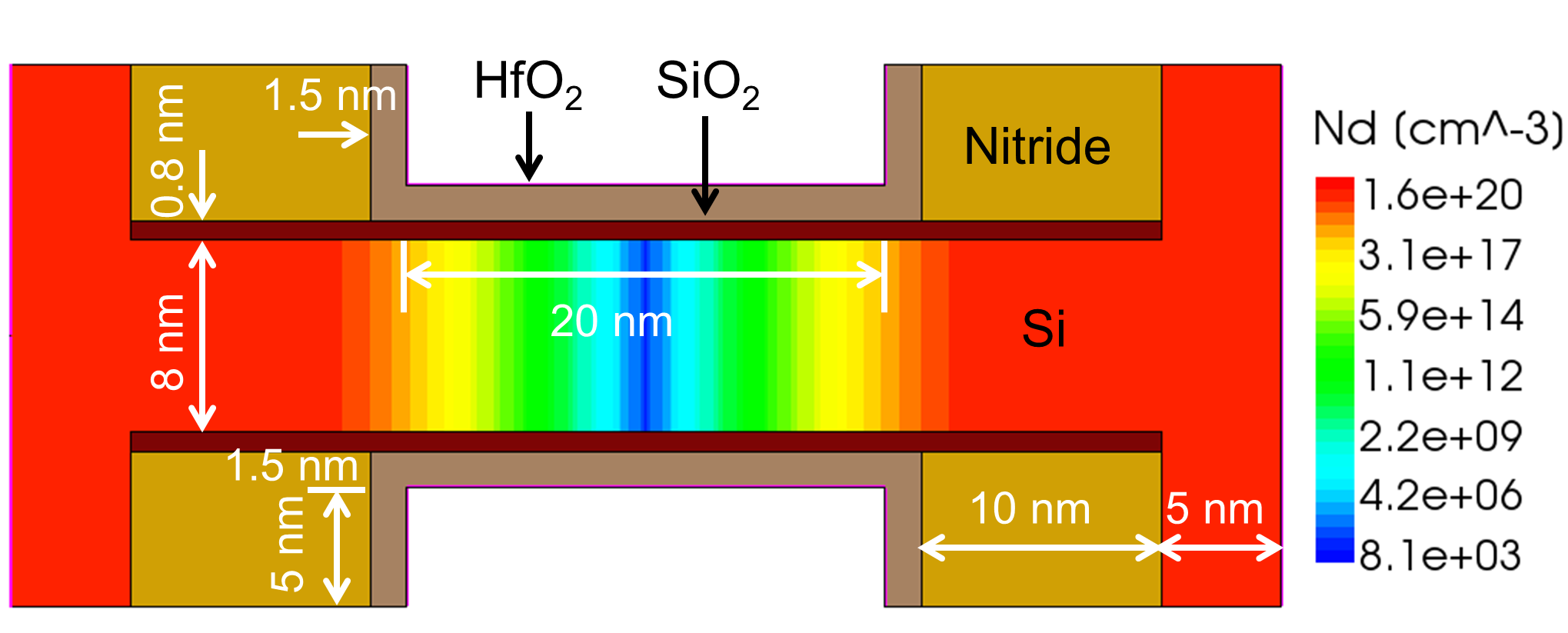}
  \caption{Simulated 2D DG-FET composed of silicon, SiO$_2$, HfO$_2$, and Nitride. Plotted donor concentration is in the log scale and a uniform acceptor concentration of $3.0\times10^{18}$ cm$^{-3}$ is used in the simulation.}
  \label{fig:full_n1}
\end{figure}

\begin{figure}[!t]
  \centering
  \includegraphics[width=3.0in]{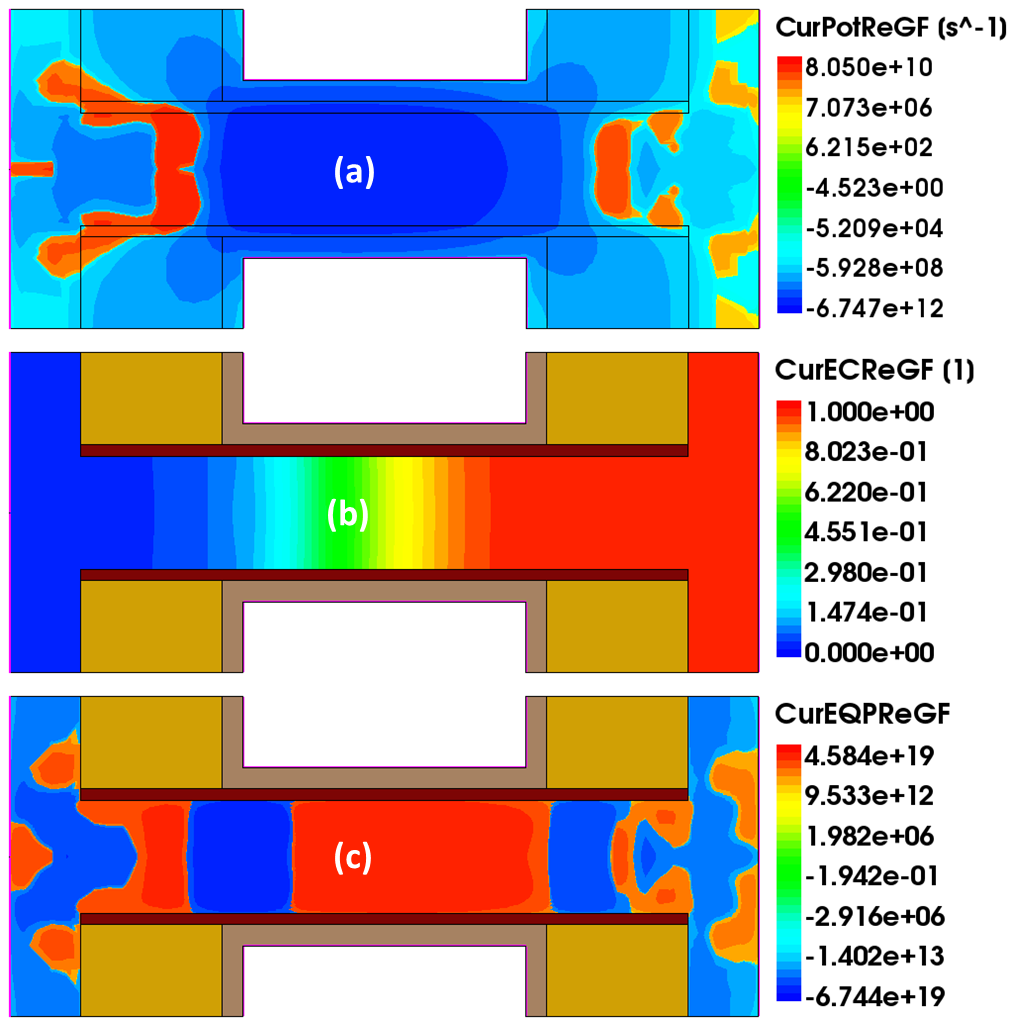}
  \caption{Current Green's functions: (a) Poisson's equation (b) electron continuity equation (c) electron density gradient equation. $V_{\text{D}}$=$V_{\text{G}}$=0.8 V.}
  \label{fig:gf_2d}
\end{figure}

Consider a double gate field effect transistor (DG-FET) as shown in Fig.~\ref{fig:full_n1}. All dimensions are marked in the plot. The current Green's functions for (\ref{eq:poissonEqu})-(\ref{eq:quanEqu_den}) are drawn in Fig.~\ref{fig:gf_2d} when both the drain and gate voltages are at 0.8 V. We will validate the GV model with numerical experiments examining silicon/oxide interface displacements and silicon body deformations in the channel region. 

\subsubsection{Interface Displacement}
\begin{figure}[!t]
  \centering
  \includegraphics[width=3.0in]{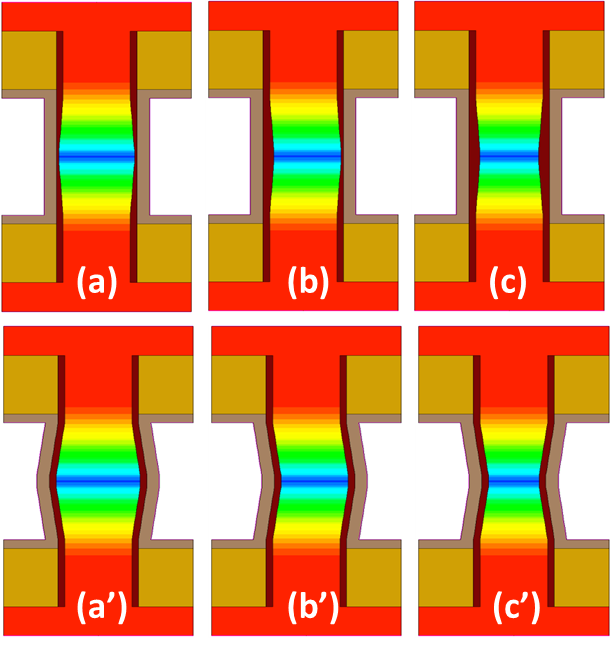}
  \caption{Illustration of geometric variation types. (a), (b), and (c) are interface displacements; (a') (b'), and (c') are body deformation. (a) and (a') are anti-correlated outward; (b) and (b') are correlated; (c) and (c') are anti-correlated inward.}
  \label{fig:full_IS_BD_shapes_all}
\end{figure}

\begin{figure}[!t]
  \centering
  \includegraphics[width=3.0in]{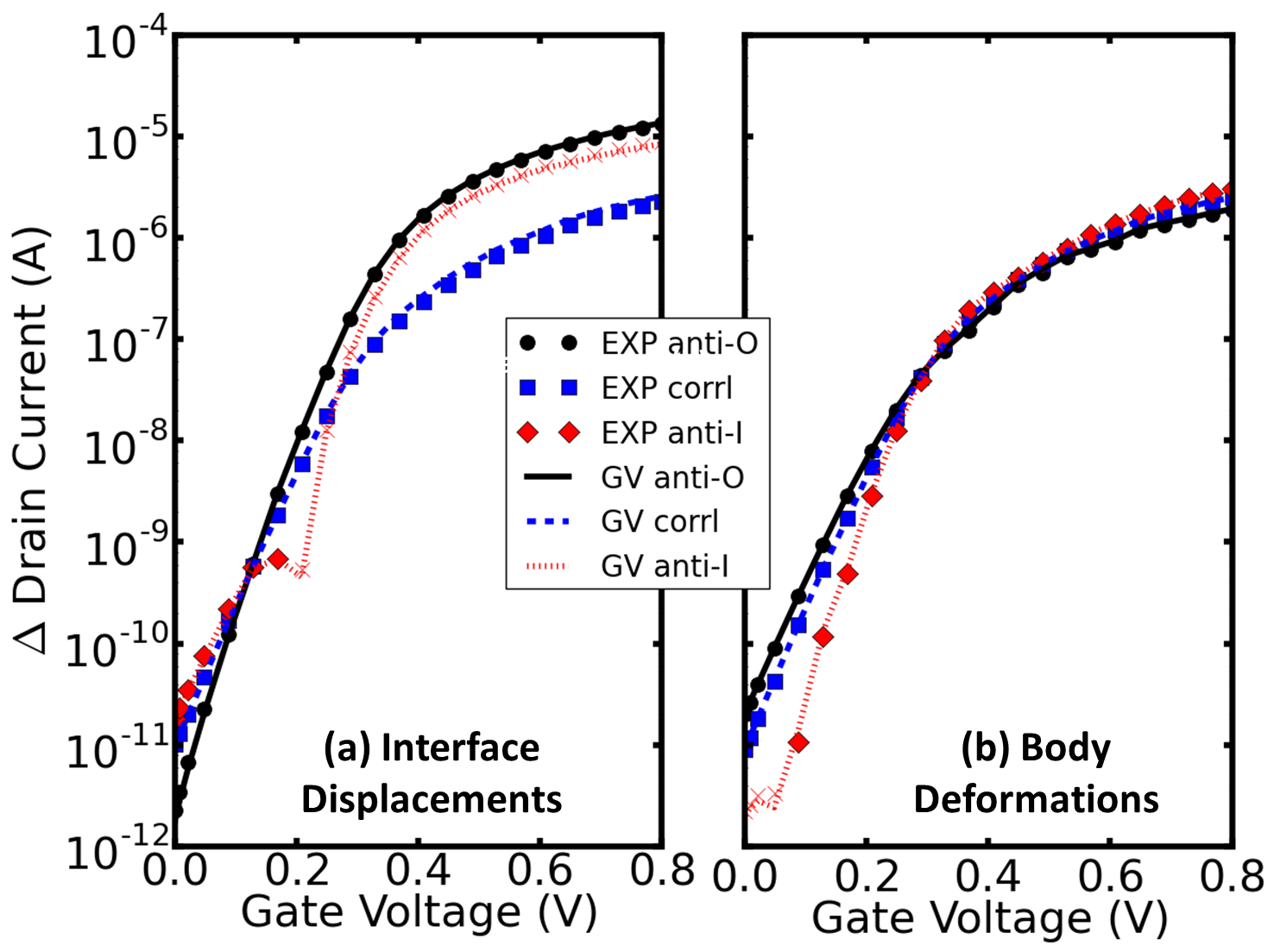}
  \caption{Comparison of drain current change from numerical experiments (markers) and GV model (lines) for (a) silicon interface displacements and (b) body deformation for Fig.~\ref{fig:full_IS_BD_shapes_all}. All values are positive except cross symbols (x) whose absolute values are used instead in the log-scale plot. $V_{\text{D}}$=0.8 V.}
  \label{fig:full_IS_BD_IV_compare}
\end{figure}

Depending on the correlation between the two gates, (a), (b), and (c) in Fig.~\ref{fig:full_IS_BD_shapes_all} show three different cases of silicon/oxide interface displacements. The maximum thickness change of the gate interfacial oxide in the middle of the channel is 0.5 nm at each side (equals to 12.5\% width variation) which is a large value for the illustration purpose. Fig.~\ref{fig:full_IS_BD_shapes_all}(a) shows an anti-correlated case, in which both silicon/oxide interfaces shift outward to make a thicker body. Fig.~\ref{fig:full_IS_BD_shapes_all}(b) has correlated variations, making the channel width constant. Fig.~\ref{fig:full_IS_BD_shapes_all}(c) also has an anti-correlated case but both interfaces shift inward, making a thinner body in the middle of channel. For each case, oxide volume change compensates for silicon volume change and vice versa, so that all of them appear to be the case of silicon/oxide interface displacements. The change of drain current by applying 1\% of geometric variation from numerical experiments (markers) and from GV (lines) are compared in Fig.~\ref{fig:full_IS_BD_IV_compare}(a). We can see that GV model's predictions agree well with the results of numerical experiments. 

$\Delta I_{\text{D}}\text{-}V_{\text{G}}$ results from GV model in Fig.~\ref{fig:full_IS_BD_IV_compare} should be combined with the $ I_{\text{D}}\text{-}V_{\text{G}}$ of the nominal device to recover the $I_{\text{D}}\text{-}V_{\text{G}}$ curves of all variational devices. Since all variations are from the same nominal device, it is convenient to evaluate the performance of variational devices on the $\Delta  I_{\text{D}}\text{-}V_{\text{G}}$ curves directly. Among the three cases, anti-correlated outward shape (black circle) corresponding to Fig.~\ref{fig:full_IS_BD_shapes_all}(a) outperforms the other two regarding both the sub-threshold slope and the on-current because thinner oxide thickness provides a better gate control. On the contrary, thicker oxide on both gates in the anti-correlated inward shape (red diamond) corresponding to Fig.~\ref{fig:full_IS_BD_shapes_all}(c) performs the worst (cross symbols stand for negative values).

\subsubsection{Body Deformation}
Three body deformation cases (a'), (b'), and (c') in Fig.~\ref{fig:full_IS_BD_shapes_all}  are derived from (a), (b), and (c) by applying the same silicon variational shapes, except keeping the oxide thickness unchanged. Fig.~\ref{fig:full_IS_BD_IV_compare}(b) compares the results of GV model (lines) with numerical experiments (markers), which also show good agreement. It also reveals that the anti-correlated inward case (red diamond) for Fig.~\ref{fig:full_IS_BD_shapes_all}(c') has the best electrostatic characteristics among the three: its on-current increases and its sub-threshold slope becomes steeper because of better gate control on the thin body.


For both silicon/oxide interface displacement and body deformation cases, GV model has shown accurate predictions confirmed by numerical experiments. Other geometric variation situations may be considered involving other body deformations which GV model can be easily applied to. Another benefit of GV model is that it can obtain terminal characteristic changes for all samples during the one nominal device simulation, while numerical experiments require running each sample separately.



\subsection{Three Dimensional FinFET}
\begin{figure}[!t]
  \centering
  \includegraphics[width=3.0in]{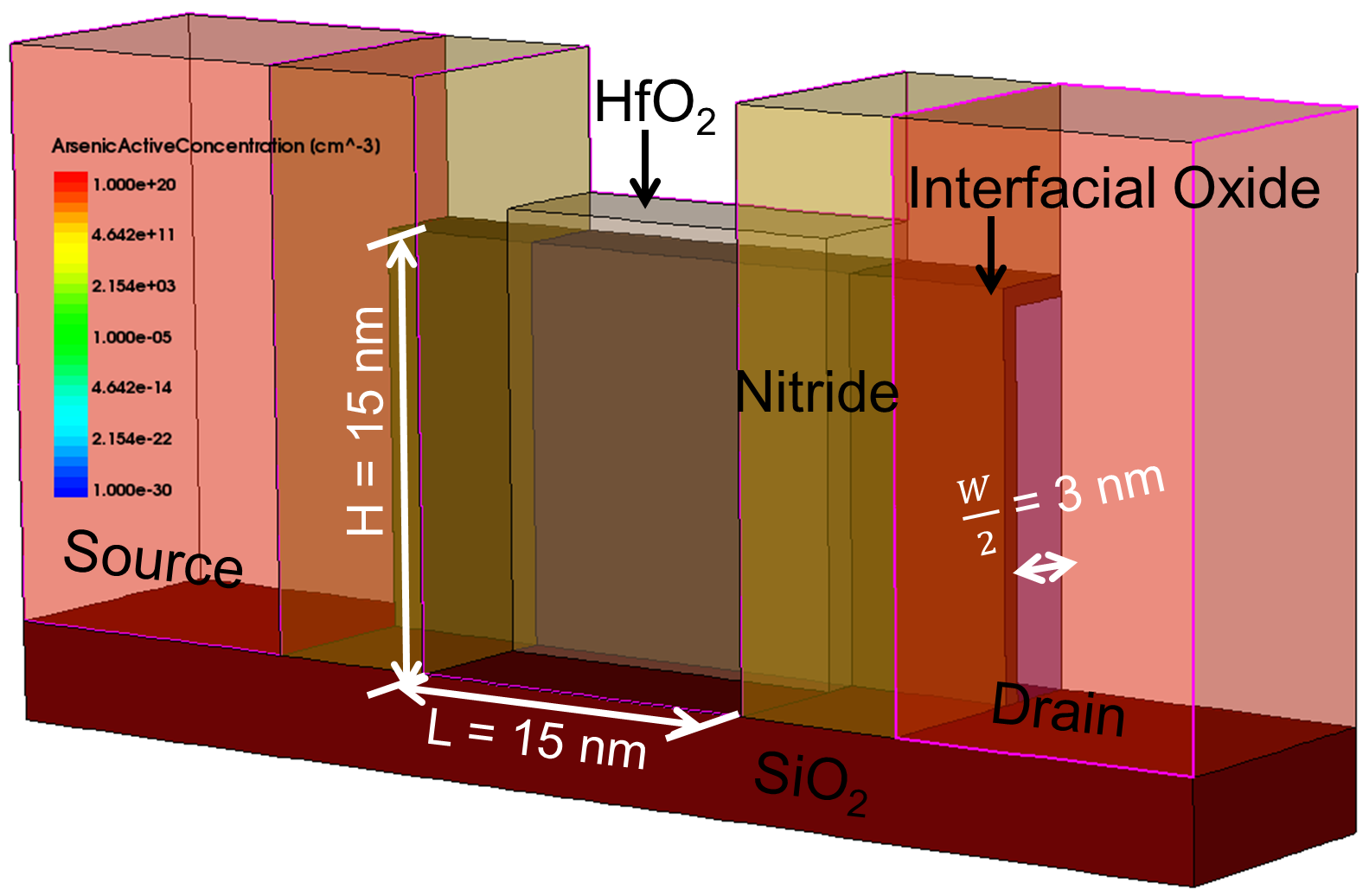}
  \caption{Three dimensional view of a half structure nMOS FinFET on a SiO$_2$ substrate. Source/drain doping is of 10$^{20}$ cm$^{-3}$ and the channel is intrinsic.}
  \label{fig:finfet_struct}
\end{figure}

For a more realistic application, we use our GV model to study the geometric variations of an n-type FinFET as shown in Fig.~\ref{fig:finfet_struct}. The FinFET has the channel length of 15 nm, the body thickness of 6 nm (half structure of 3 nm is shown), and the fin height of 15 nm. In the following simulations, we keep drain voltage at 0.8V. Gate voltage increases from 0 to 0.8V.

\subsubsection{Linearity}
\begin{figure}[!t]
  \centering
  \includegraphics[width=3.0in]{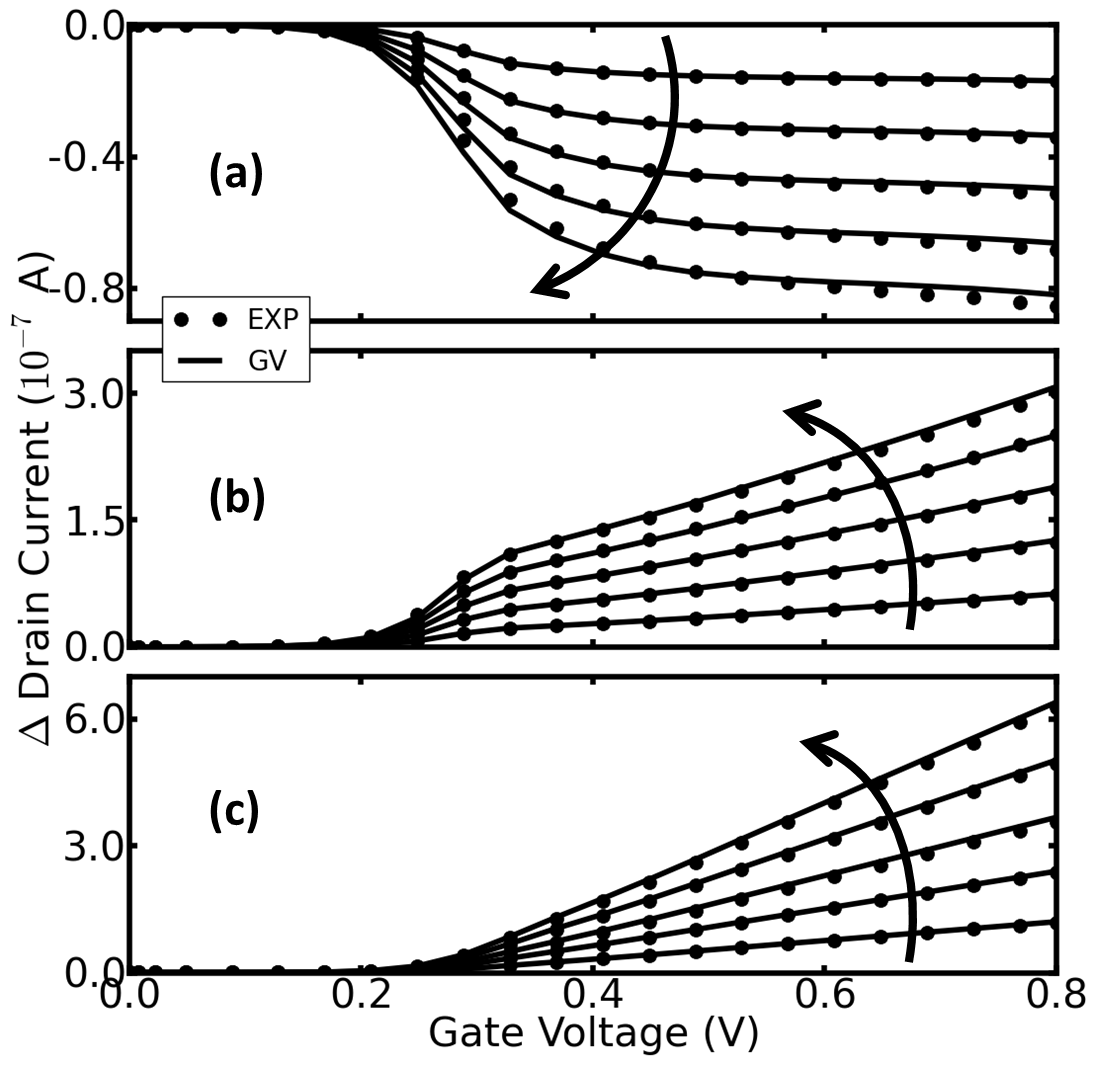}
  \caption{$\Delta I_{\text{D}}\text{-}V_{\text{G}}$ curves with 2\% increment of geometric variation from 2\% to 10\% on (a) length, (b) width, and (c) height. Arrows point to the increase of variations.}
  \label{fig:finfet_linear}
\end{figure}

Multiple variation amplitudes from 2\% to 10\% with a steady step of 2\% have been applied uniformly on the device channel length, body width, and fin height. Corresponding drain current changes are calculated using the GV model, which are compared with numerical experiments in Fig.~\ref{fig:finfet_linear}. Generally, GV results match with numerical experiments and the accuracy decreases with the increase of noise source magnitude. Comparing the absolute values of $\Delta I_{\text{D}}$ in the three cases, Fig.~\ref{fig:finfet_linear} also reveals that given the same geometrical variation ratio, the on-current of the FinFET in Fig.~\ref{fig:finfet_struct} is most sensitive to the height variation (Fig.~\ref{fig:finfet_linear}(c)), followed by the width (Fig.~\ref{fig:finfet_linear}(b)), and least sensitive to the length variation (Fig.~\ref{fig:finfet_linear}(a)). With an increase of channel length, it is shown that drain current decreases (negative $\Delta$Id). In comparison, the expanded cross section by the increase of the body width or fin height leads to the increase of the drain current.


\subsubsection{Effective Noise Source}
\begin{figure}[!t]
  \centering
  \includegraphics[width=3.0in]{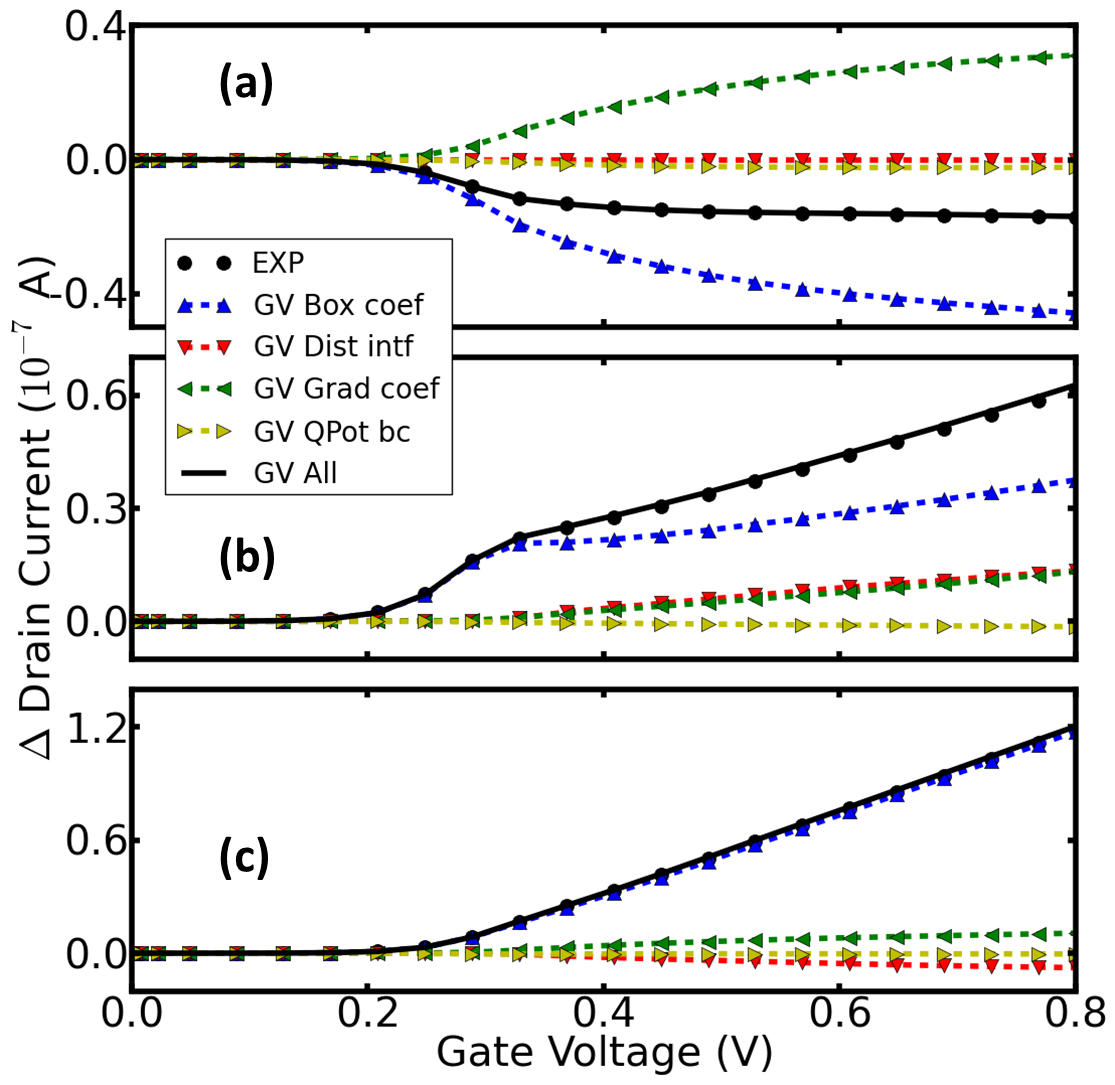}
  \caption{Contributions of effective noise source components on drain current change for 2\% geometric variations of (a) length, (b) width, and (c) height.}
  \label{fig:finfet_model}
\end{figure}

As discussed in the effective noise source calculation, calculating $\delta \mathcal{N}$ has to take into account not only the box coefficient and volume changes, but also the change of flux and source terms in (\ref{eq:pde_int}). For the physical models we have applied, four components are considered for the geometric variations: box coefficient $A$ and volume $\Omega$ (Box coef), node distance to the interface (Dist intf), gradient of box coefficient (Grad coef), and quantum potential boundary condition (QPot bc). Fig.~\ref{fig:finfet_model} shows the each noise source contribution by the 2\% variation of length, width, and height. Only considering box coefficient and volume (Box coef), GV's predictions can be quite off in the case of length and width variations. In the length variation (Fig.~\ref{fig:finfet_model}(a)), node distance to interfaces does not change so that it (Dist intf) gives a zero influence on the total result. Fig.~\ref{fig:finfet_model}(b) suggests that Box coef, Dist intf, and Grad coef contribute to the same direction in the width variation. Height variation is dominated by the box coefficient and volume (Box coef), and net effect of others only provides minor corrections.

\subsubsection{Equation Contribution}

\begin{figure}[!t]
  \centering
  \includegraphics[width=3.0in]{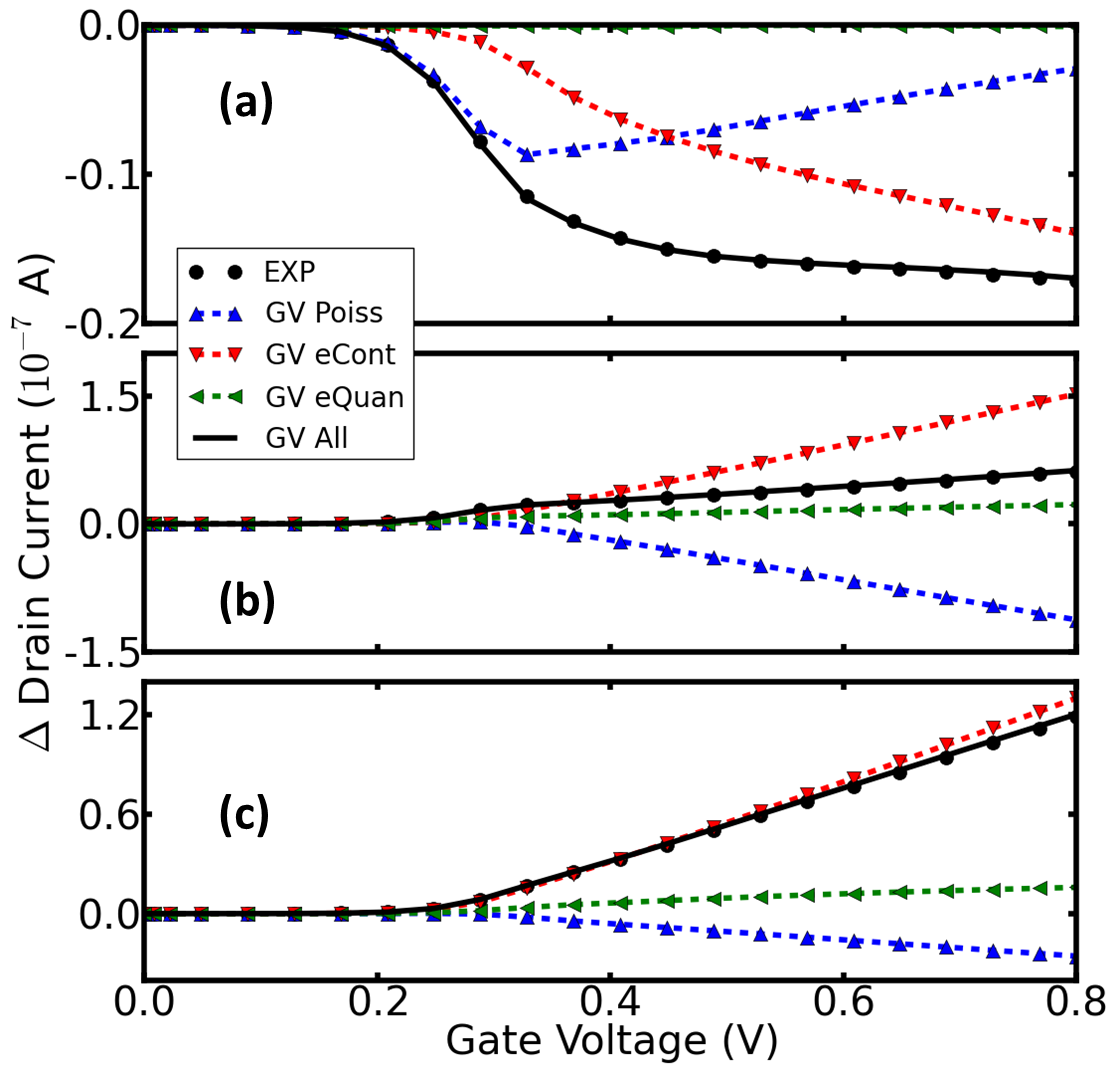}
  \caption{Individual equation and their total contribution on drain current change for 2\% geometric variations of (a) length, (b) width, and (c) height. For nMOS, hole contributions from the continuity and density-gradient quantum potential equations are negligible for all cases so they are not shown. }
  \label{fig:finfet_compare_each}
\end{figure}

For the case of the 2\% variation, a detailed examination of the contribution from each equation in (\ref{eq:poissonEqu})-(\ref{eq:quanEqu_den}) for the three cases is illustrated in Fig.~\ref{fig:finfet_compare_each}. In the length variation (Fig.~\ref{fig:finfet_compare_each}(a)), the electron continuity and Poisson's equations are two main contributing sources, and both of them decrease the drain current. Increasing either body width (Fig.~\ref{fig:finfet_compare_each}(b)) or fin height (Fig.~\ref{fig:finfet_compare_each}(c)) enlarges the cross section, so that the electron continuity equation brings up the drain current. An enlarged cross section also reduces the quantum confinement, thus electron density-gradient potential equation contributes a small positive amount. The negative contribution of Poisson's equation in the width case (Fig.~\ref{fig:finfet_compare_each}(b)) cancels out a significant part of the total drain current change, while in the height variation (Fig.~\ref{fig:finfet_compare_each}(c)) the effect of Poisson's equation is small, only comparable to the density-gradient equation's contribution, leaving the net drain current change larger in the height variation case.




\subsubsection{Sensitivity Map}
\begin{figure}[!t]
\centering
\subfloat[Poiss]{
  \includegraphics[width=1.0in, height=1.5in]{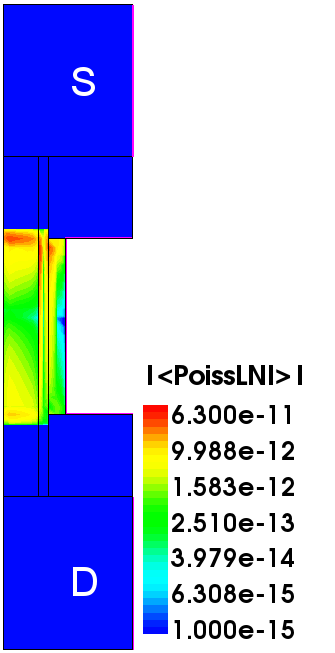}%
  \label{fig:finfet_length_sen_poiss}
}
\hfil
\subfloat[eCont]{
  \includegraphics[width=1.0in, height=1.5in]{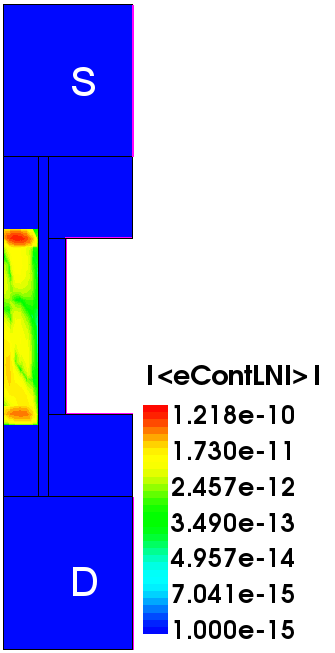}%
  \label{fig:finfet_length_sen_eCont}
}
\hfil
\subfloat[eQuan]{
  \includegraphics[width=1.0in, height=1.5in]{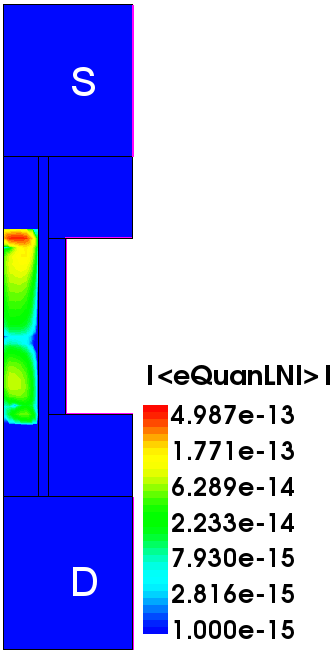}%
  \label{fig:finfet_length_sen_eQuan}
}
\caption{Sensitivity maps of equation contributions of drain current local noise for 2\% geometric variations of channel length.}
\label{fig:finfet_length_sen}
\end{figure}

\begin{figure}[!t]
\centering
\subfloat[Poiss]{
  \includegraphics[width=1.0in, height=1.5in]{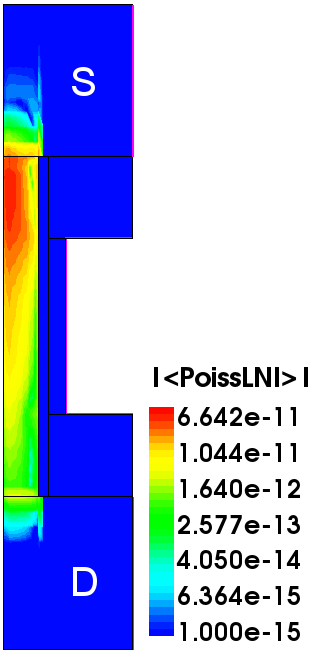}%
  \label{fig:finfet_width_sen_poiss}
}
\hfil
\subfloat[eCont]{
  \includegraphics[width=1.0in, height=1.5in]{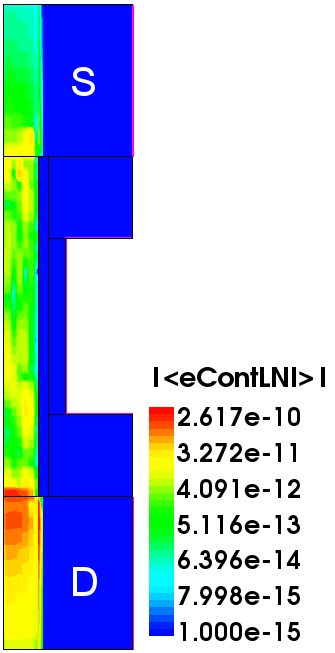}%
  \label{fig:finfet_width_sen_eCont}
}
\hfil
\subfloat[eQuan]{
  \includegraphics[width=1.0in, height=1.5in]{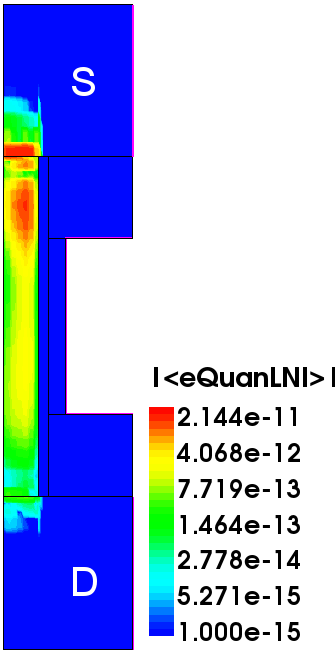}%
  \label{fig:finfet_width_sen_eQuan}
}
\caption{Sensitivity maps of equation contributions of drain current local noise for 2\% geometric variations of body width.}
\label{fig:finfet_width_sen}
\end{figure}

\begin{figure}[!t]
\centering
\subfloat[Poiss]{
  \includegraphics[width=2.5in, height=1.0in]{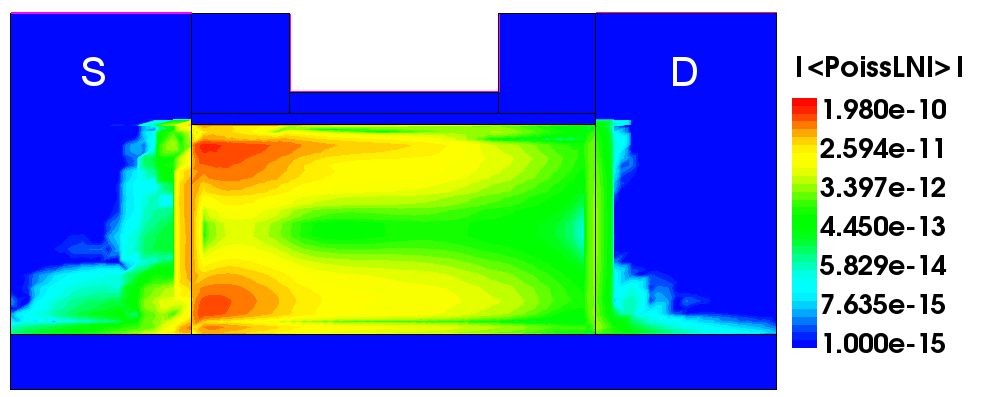}%
  \label{fig:finfet_height_sen_poiss}
}
\hfil
\subfloat[eCont]{
  \includegraphics[width=2.5in, height=1.0in]{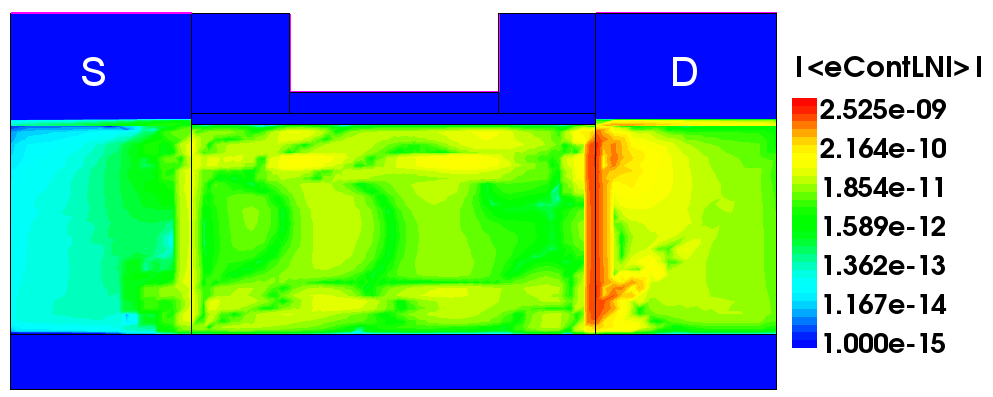}%
  \label{fig:finfet_height_sen_eCont}
}
\hfil
\subfloat[eQuan]{
  \includegraphics[width=2.5in, height=1.0in]{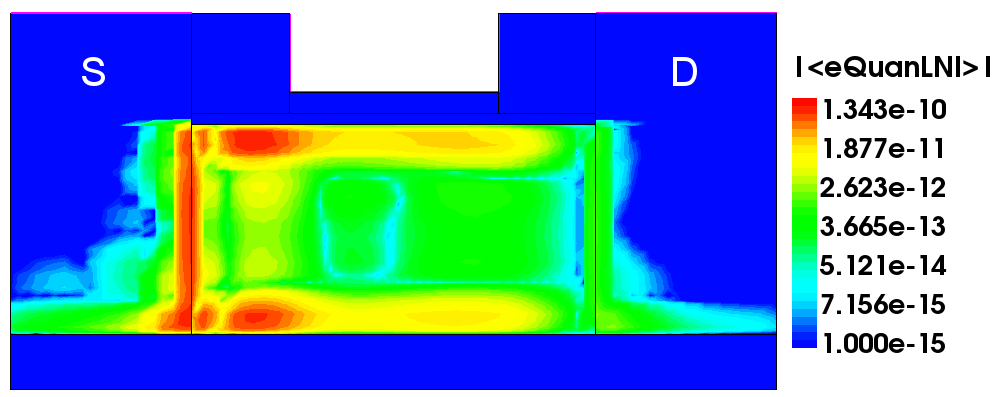}%
  \label{fig:finfet_height_sen_eQuan}
}
\caption{Sensitivity maps of equation contributions of drain current local noise for 2\% geometric variations of fin height.}
\label{fig:finfet_height_sen}
\end{figure}

The last benefit of applying GV model is that geometrically sensitive areas which mostly influence terminal characteristics can be easily identified. At both drain and gate voltages equal to 0.8V, for the three variational cases we can plot sensitivity maps. Fig.~\ref{fig:finfet_length_sen} shows a plane along transport direction close to the SiO$_2$ substrate. The plots show the node-wise absolute value of local noise of the drain current in the log scale for a uniform 2\% channel length variation. We can see that the electron density gradient equation gives contribution (Fig.~\ref{fig:finfet_length_sen}(c)) of two orders magnitude smaller than the Poisson's and electron continuity equations. The largest drain current local spatial noise for Poisson's equation comes from the source/channel boundary where we have applied geometric variation and also the channel/oxide interface close to the source side (Fig.~\ref{fig:finfet_length_sen}(a)). For the electron continuity equation, both source/channel and drain/channel boundaries are important although source side is contributing more (Fig.~\ref{fig:finfet_length_sen}(b)). Similarly for the body width variation, sensitivity maps are drawn in Fig.~\ref{fig:finfet_width_sen}. Source/channel overlap region is important for the Poisson's equation (Fig.~\ref{fig:finfet_width_sen}(a)). A small part of drain region which is adjacent to the channel shows high noise from the electron continuity equation (Fig.~\ref{fig:finfet_width_sen}(b)). For height variation, Fig.~\ref{fig:finfet_height_sen}(b) indicates that the leading contribution from the electron continuity equation comes from the drain/channel interface. Poisson's equation's contribution mainly comes from the silicon/oxide corners on the source side (Fig.~\ref{fig:finfet_height_sen}(a)), which is similar to the case of the electron density gradient equation (Fig.~\ref{fig:finfet_height_sen}(c)). The quality of source/channel interface also has a big impact on the noise contribution from the electron density gradient equation (Fig.~\ref{fig:finfet_height_sen}(c)). One should note that because nonlocal effects (distance to interface, mesh gradient, etc.) are considered in the local noise source calculation, the plotted local contribution in Figs.~\ref{fig:finfet_length_sen}-\ref{fig:finfet_height_sen} appears to be somewhat noisy especially for the most affected electron continuity equation case. In all, Poisson's and electron density gradient equations are more sensitive to the source/channel interface, while electron continuity equation depends on the drain/channel interface.

\section{Conclusions}
To accurately and efficiently address the increasing importance of geometric variations in nanoscale electronic devices, we have proposed the GV model. As a comparison, we have explored physical limitations of the existing RGF model regarding interface displacement and body deformation. The fundamental difference between the proposed GV model and RGF model is on the noise source origins. GV considers the geometric variation directly by mesh deformation, and computes the corresponding noise sources consistently with the mesh discretization and the employed physical models. On the contrary, RGF model relies only on the interface quantities to compute the effective noise source. As a result, the RGF model cannot capture the important nonlocal geometric dependences. GV, on the other hand, relies on the Green's functions of coupled equations which are intrinsically globally defined, and thus applicable to wider variety of geometric variations.

Regardless of the interface displacement or body deformation, by taking into account all effective noise source components: box coefficient and volume, distance to the interface, gradient of box coefficient, and proper boundary condition, GV model can accurately predict the changes of terminal characteristics, which have been extensively validated by comparing with the results from numerical experiments of full device simulations.

GV model also offers great details for internal quantities: contributions of effective noise source components can be compared quantitatively, equation-wise noise contributions can be also mutually compared, and node-wise geometric sensitivity maps can be used for identifying sensitive regions. 

GV model can be a useful tool for geometric variability analysis.



\ifCLASSOPTIONcaptionsoff
  \newpage
\fi



\bibliographystyle{IEEEtran}
\bibliography{ref}
\end{document}